# Omni-resonant optical micro-cavity

Soroush Shabahang[1], H. Esat Kondakci[1], Massimo L. Villinger[1], Joshua D. Perlstein[2], Ahmed El Halawany[1], and Ayman F. Abouraddy[1,2*]

[1]CREOL, The College of Optics & Photonics, University of Central Florida, Orlando, FL 32816, USA

[2]Materials Science and Engineering Department, College of Engineering and Computer Science, University of Central Florida, Orlando, FL 32816, USA

*raddy@creol.ucf.edu

**Optical cavities are a cornerstone of photonics[1]. They are indispensable in lasers, optical filters, optical combs[2] and clocks[3], in quantum physics[4], and have enabled the detection of gravitational waves[5]. Cavities transmit light only at discrete resonant frequencies, which are well-separated in micro-structures. Despite attempts at the construction of 'white-light cavities'[6-9], the benefits accrued upon optically interacting with a cavity – such as resonant field buildup – have remained confined to narrow linewidths. Here, we demonstrate achromatic optical transmission through a planar Fabry-Pérot micro-cavity via angularly multiplexed phase-matching that exploits a bio-inspired grating configuration[10,11]. By correlating each wavelength with an appropriate angle of incidence, a continuous spectrum resonates and the micro-cavity is rendered transparent. The locus of a single-order 0.7-nm-wide resonance is de-slanted in spectral-angular space to become a 60-nm-wide achromatic resonance spanning multiple cavity free-spectral-ranges. This approach severs the link between the resonance bandwidth and the cavity-photon lifetime, thereby promising resonant enhancement of linear and nonlinear optical effects over broad bandwidths in ultrathin devices.**

Optical-cavity resonances occupy narrow spectral linewidths that are inversely proportional to the cavity-photon lifetime, which are separated by a free spectral range (FSR) that is inversely proportional to the cavity size. Although cavity-quantum electrodynamics requires narrow cavity linewidths to isolating the interaction of optical fields with the resonances of atoms, ions, or nanostructures[4], most applications would benefit from maintaining the resonant cavity field buildup over an extended bandwidth. Examples of such applications include coherent perfect absorption (CPA) in media endowed with low intrinsic losses[12,13] and boosting nonlinear optical effects[14]. Although CPA, for instance, can increase absorption to 100% in a thin low-loss layer *on* resonance[15], exploiting CPA in harvesting solar radiation would require an optical cavity in which an *extended bandwidth* satisfies the resonance condition.

The quest for producing an achromatic resonator has precedents. In 'white-light cavities', the cavity itself is modified by inserting a new material or structure endowed with strong negative (anomalous) dispersion to equalize the cavity optical length for all wavelengths[6]. Only macroscopic white-light cavities have been explored to date via cavity-filling atomic species featuring bifrequency Raman gain in a double-Λ system[7] or displaying electromagnetically induced transparency[8], or alternatively via nonlinear Brillouin scattering[9]. In all such studies, the enhanced cavity linewidths are extremely narrow (~ 100 MHz or < 1-pm-wide) by virtue of the very nature of the atomic or nonlinear resonances utilized, and are limited by uncompensated higher-order dispersion terms. Alternative approaches based on the use of *linear* optical components, such as appropriately designed chirped cavity mirrors[16] or grating pairs[17], have been investigated. Surprisingly, both of these possibilities fail at producing a white-light cavity due to subtle overlooked aspects in the constraints imposed by causality on non-dissipative systems[18,19].

Here we demonstrate achromatic transmission through a planar Fabry-Perot micro-cavity – *not* by modifying its structure, but instead by altering the spectral-spatial configuration of the incident optical radiation using linear optical components. We show that the spatial degree-of-freedom of the optical field when used *in conjunction* with its spectral degree-of-freedom altogether obviates the limitations inherent in traditional approaches to constructing a white-light cavity. In place of narrow well-separated resonant linewidths of a micro-cavity, broadband 'achromatic resonances' emerge. Starting from the curved locus of a cavity resonance in spectral-angular space, we *de-slant* this locus through angular multiplexing of incident broadband light. Achromaticity is achieved by establishing a judicious correlation between the wavelengths and their associated incident angles, which results in optical 'clearing' of the cavity. Anomalous angular diffraction – achieved via a bio-inspired grating configuration[10,11] – engenders the necessary correlation and enables continuous phase-matching of the wave-vector axial component to fulfil the resonance condition over an extended bandwidth. We demonstrate this effect using a planar micro-cavity whose linewidth is ~ 0.7-nm-wide and FSR is ~ 25 nm. Single-order ~ 60-nm-wide resonances that span multiple original FSRs emerge, thereby rendering the resonator transparent – and even enabling the formation of an image through it. In principle, such achromatic resonances can be established over an indefinitely wide bandwidth by replacing the grating with an appropriately designed metasurface[20].

The underlying physical principle for realizing achromatic resonances in a planar Fabry-Pérot cavity can be understood by referring to Fig. 1. At normal incidence (Fig. 1a), only discrete wavelengths resonate whose associated roundtrip phase $\varphi$ is an integer multiple of $2\pi$, $\varphi(\lambda) = 2nkd + 2\gamma(\lambda) = 2\pi m$; here $\lambda$ is the free-space wavelength, $k = \frac{2\pi}{\lambda}$ is the wave number, $d$ and $n$ are the thickness and refractive index of the cavity layer, respectively, integer $m$ is the resonant-mode order, and $\gamma$ is the reflection phase from the cavity mirrors[1] (assumed symmetric). At an incidence angle $\theta$, the resonances are *blue-shifted* (Fig. 1b) because only the *axial* component of the wave vector contributes to the phase $\varphi(\lambda, \theta) = 2nkd\cos\theta' + 2\gamma(\lambda, \theta') = 2\pi m$, where $\theta'$ is the angle inside the cavity corresponding to an external angle $\theta$ through Snell's law. Indeed, for every wavelength $\lambda$, there is an angle $\theta(\lambda)$ that enables this particular wavelength to resonate by satisfying the phase-matching condition

$$\varphi(\lambda,\theta) = 2nkd\cos[\theta'(\lambda)] + 2\gamma(\lambda,\theta') = 2\pi m. \qquad (1)$$

Therefore, by re-organizing the incident broadband radiation by assigning each wavelength $\lambda$ to an appropriate incidence angle $\theta(\lambda)$, all the angularly multiplexed wavelengths can resonate simultaneously (Fig. 1c), with shorter wavelengths requiring larger incidence angles. Hence, by providing a pre-compensation tilt angle to each wavelength prior to incidence, such that $k\cos[\theta'(\lambda)]$ is constant, we effectively de-slant the resonance by maintaining $\varphi(\lambda,\theta)$ independent of $\lambda$ (the horizontal line in Fig. 1d).

We first present a heuristic argument for the construction of an optical system that de-slants a resonance in spectral-angular space (Fig. 2a). A 'black box' system that implements any of the targeted correlations $\theta(\lambda)$ shown in Fig. 2b will enable a broadband beam to transmit through the cavity via angular multiplexing – with all the wavelengths resonating simultaneously – and then its inverse restores the original beam. Dispersive prisms do not provide the required angular spread, and planar surface gratings produce the opposite correlation: longer wavelengths diffract at larger angles with respect to the normal as a consequence of transverse phase-matching (dashed curve in Fig. 2b)[21]. In other words, the spatial-spectral dispersion inculcated by an optical grating and by a cavity – two of the most fundamental optical devices – are in opposition. Instead, so-called 'anomalous diffraction' or 'reverse-color sequence' is required. To address this challenge, we take our inspiration from the reverse-color sequence observed in the diffraction of white light off the wing scales of the butterfly *Pierella luna*[10]. This effect has been revealed to be *geometric* in nature: 'vertical' micro-gratings that grow on the *Pierella luna* scales reverse the sequence of diffracted colors as confirmed by fabricated artificial counterparts[11]. We adopt this strategy here in reflection mode and vary the relative tilt between the grating and the cavity, from 0° in Fig. 2c to 90° in Fig. 2d, to enable a transition from normal to anomalous diffraction, respectively.

To gain insight into the resonance de-slanting procedure, we first examine the spectral-angular variation in the axial wave-vector component $k_z$ of broadband light propagating in a 'bulk' planar layer of refractive index $n$. Consider a bandwidth $\Delta\lambda$ centered at $\lambda_c$ and each wavelength is directed at a different angle $\theta(\lambda)$, with $\theta(\lambda_c) = \psi$, such that the beam occupies an angular spread $\Delta\theta$ (assume the wavelengths are distributed uniformly around $\psi$). For a wavelength $\lambda$ incident at an external angle $\theta$, $k_z$ in the layer is

$$k_z(\lambda,\psi;\beta) = \frac{2\pi}{\lambda}\sqrt{n^2 - \sin^2[\psi - \beta(\lambda - \lambda_c)]}, \qquad (2)$$

where $\beta = \Delta\theta/\Delta\lambda$ °/nm is the angular dispersion, we take $n = 1.5$ and $\lambda_c = 550$ nm, and we ignore the spectral variation of $n$ for simplicity. We search for a region in $(\lambda,\psi)$ space where $k_z$ is *independent* of $\lambda$. We plot in Fig. 2e the value of $k_z$ for several values of angular dispersion $\beta$. When mirrors sandwich a layer of thickness $d$, resonances are established whenever $k_z$ is an integer multiple of $\frac{\pi}{d}$. Setting $\beta = 0$, we retrieve the case of collimated light incident on a planar layer at an external angle of incidence $\psi$. As $\beta$ increases, the constant-$k_z$ contours display less curvature with respect to $\lambda$. At $\beta = 0.37$ °/nm we reach a critical condition where $k_z$ over an extended region in $(\lambda,\psi)$ space becomes *independent* of $\lambda$. A broadband optical beam prepared in this configuration will transmit through a cavity via achromatic resonances supported in this region. Increasing $\beta$ further reverses the curvature of the constant-$k_z$ contours with respect to $\lambda$, thereby disrupting the achromatic resonances.

We have carried out an experiment to confirm this prediction of achromatic resonances utilizing a Fabry-Pérot cavity consisting of a 4-μm-thick layer of $SiO_2$ ($n = 1.48$ at $\lambda = 550$ nm) sandwiched between two Bragg mirrors each formed of 5 bilayers. Each bilayer comprises 92.2-nm and 65.5-nm-thick layers created by the evaporation of SiO$_2$ and Ti$_2$O$_3$ ($n = 2.09$ at $\lambda = 550$ nm), respectively, to produce a 120-nm-wide reflection band with ~ 92% reflectivity at its center wavelength $\lambda_c \approx 550$ nm at normal incidence. The cavity (total thickness $\approx 5.6$ µm) is deposited monolithically by electron-beam evaporation on a 0.5-mm-thick, 25-mm-diameter glass slide (Fig. 3b, inset). Figure 3a depicts the measured spectral-angular transmission through the cavity obtained using a ~ 3-mm diameter collimated white-light beam from a halogen lamp revealing the standard behavior of a planar micro-cavity[22]. Upon normal incidence, a finite set of resonant wavelengths are transmitted with a FSR of $\approx 25$ nm, which are blue-shifted with angle of incidence $\theta$; see Supplementary for details.

We next modify the collimated white-light beam to produce the necessary condition to de-slant the resonance locus – without altering the cavity itself in any way. The beam is first spatially filtered through a 1-mm-wide vertical slit (to avoid aliasing of multiple resonance orders) and is then diffracted from a reflective grating with 1800 lines/mm (Fig. 3b). The grating produces an angular dispersion of $\beta \approx 0.09$ °/nm at $\lambda_c = 550$ nm. A grating with ~ 3500 lines/mm produces the target $\beta$, but such a high-density grating has a low diffraction-efficiency in the visible. To enhance $\beta$, we add a lens in the path of the diffracted beam before the cavity (L$_1$ in Fig. 3b). The spectral transmission through the Fabry-Pérot cavity with tilt angle $\psi$ is plotted in Fig. 3c,d. It is critical to note that the angle $\psi$ is *not* the incidence angle of the beam onto the cavity, but is instead simply the tilt angle of the cavity with respect to the central wavelength $\lambda_c = 550$ nm that defines the optical axis (see Supplementary); each wavelength is in fact incident at its own angle $\theta(\lambda)$. For convenience, we hold the grating fixed and rotate the cavity. Using L$_1$ with focal length $f = 50$ mm, $\beta$ is enhanced to 0.13 °/nm, and the blue-shift of the resonance loci is boosted (Fig. 3c). Reducing the focal length of L$_1$ to $f = 25$ mm increases $\beta$ further and reaches the desired angular/spectral dispersion (corresponding to the third panel in Fig. 2e). The resonance loci are now flattened horizontally at specific values of $\psi$, whereupon all the wavelengths extending across a 60-nm-wide bandwidth – exceeding twice the FSR – resonate simultaneously (Fig. 3d), a phenomenon we name achromatic resonance.

As a result of the cavity achromaticity, one may indeed image an object through the cavity with broadband illumination. We add a lens to the setup in Fig. 3b to image a plane preceding the grating to a plane lying beyond the cavity (see Supplementary). The object is a binary-valued 0.25×2 mm$^2$ transparency of the letter 'i' that is imaged through the cavity with a magnification factor of ~ 3. In absence of the grating, a limited amount of light is transmitted through the cavity at any incident angle due to the large FSR and narrow linewidth of the resonances lying within the cavity mirror bandgap (Fig. 4a,b) – when compared to the configuration where the cavity is absent (Fig. 4c). In presence of the grating that renders the cavity transparent, a substantial amount of light is transmitted when the cavity tilt angle corresponds to that of an achromatic resonance: at $\psi = 30°$, 39°, 48°, and 57° (Fig. 4a,d).

Our proof-of-principle experiment renders transparent a micro-cavity with 0.7-nm-wide resonances separated by an FSR of ~ 25 nm, thanks to an achromatic resonance operating continuously over a broad spectrum (~ 60 nm). Although the necessary correlation between wavelength and incidence angle is introduced using a planar surface grating, the bandwidth can be broadened further and the uniformity of the spectral transmission improved by replacing the grating with a metasurface realizing a customized function $\theta(\lambda)$ that takes into account the cavity mirror spectral phase $\gamma(\lambda, \theta)$, its polarization dependence, and wavelength dependence of the refractive index[23]. Furthermore, such a metasurface may indeed implement the reverse-color sequence without introducing a tilt angle with respect to the cavity[24]. Consequently, depositing the metasurface directly on the planar micro-cavity may potentially result in ultra-thin optical devices that deliver resonant linear and nonlinear behavior over extended bandwidths.

We have introduced here a general principle that lifts the bandwidth restrictions associated with resonant linewidths in an optical micro-cavity – leading to the realization of an achromatic or white-light cavity. While recent work has exploited spectral splitting of the solar spectrum to optimize the photovoltaic conversion with multiple semiconductor junctions[25], our approach – on the other hand – implements a continuous mapping to a wavelength-dependent angle of incidence $\theta(\lambda)$. Indeed, our work extends to the continuum the correlations between discretized optical degrees of freedom studied in Refs. 26, 27. As a result, the advantages associated with a resonance – such as field enhancement through resonant buildup and enhanced optical nonlinearities – become altogether decoupled from the cavity linewidth and are thus available over orders-of-magnitude larger bandwidths. This concept can have a profound impact on optics by bringing coherent perfect absorption to bear on harvesting solar energy, producing white-light micro-lasers, and yielding broadband resonantly enhanced nonlinear optical devices.

**Acknowledgments.** We thank Boris Y. Zeldovich, Leonid B. Glebov, and Michael P. Marquez for helpful discussions. This work was supported by the US Air Force Office of Scientific Research (AFOSR) through MURI award FA9550-14-1-0037 and the US Office of Naval Research (ONR) through award N00014-14-1-0260.

**Figures and captions**

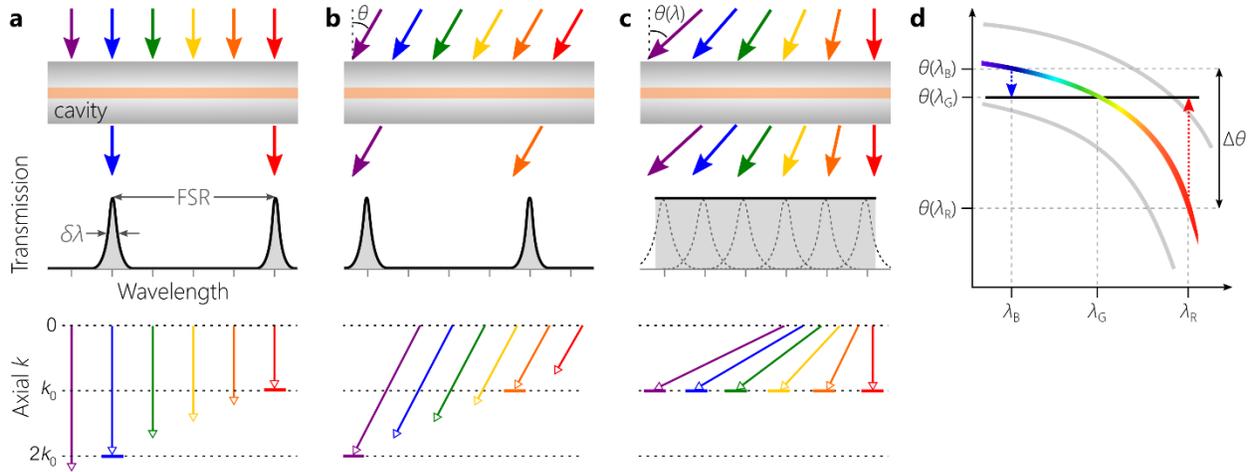

**Figure 1 | Spectral-angular correlations produce achromatic resonances in a micro-cavity. a**, When collimated broadband light is incident normally on a planar Fabry-Pérot cavity (top row), only a discrete set of wavelengths transmit (middle row) whose axial component of the wave vector inside the cavity is an integer multiple of $k_\text{o} = \frac{\pi}{d}$ (identified by solid horizontal dashes in the bottom row); $\delta\lambda$ is the resonance linewidth. **b**, The cavity resonances are blue-shifted when light is incident at an angle $\theta$. **c**, By assigning each wavelength $\lambda$ to an appropriate angle of incidence $\theta(\lambda)$, all the wavelengths can resonate and transmission becomes achromatic. One resonant order can extend here over a bandwidth exceeding the FSR. **d**, Locus of resonant orders in spectral-angular space. Fixing the angle of one wavelength $\theta(\lambda_\text{G})$, we can de-slant the resonance of a specific order (colored curve) by boosting and reducing a pre-compensation angle for each wavelength to produce an achromatic resonance (solid horizontal line). An angular spread $\Delta\theta$ at the input is required to de-slant the resonance between $\lambda_\text{B}$ and $\lambda_\text{R}$. At the shorter wavelength $\lambda_\text{B}$, the incidence angle needs to be increased above $\theta(\lambda_\text{G})$ by $\theta(\lambda_\text{B}) - \theta(\lambda_\text{G})$. The longer wavelength $\lambda_\text{R}$ requires an incidence angle lower than $\theta(\lambda_\text{G})$ by $\theta(\lambda_\text{G}) - \theta(\lambda_\text{R})$. Consequently $\lambda_\text{R}$, $\lambda_\text{G}$, and $\lambda_\text{B}$ all satisfy the resonance condition.

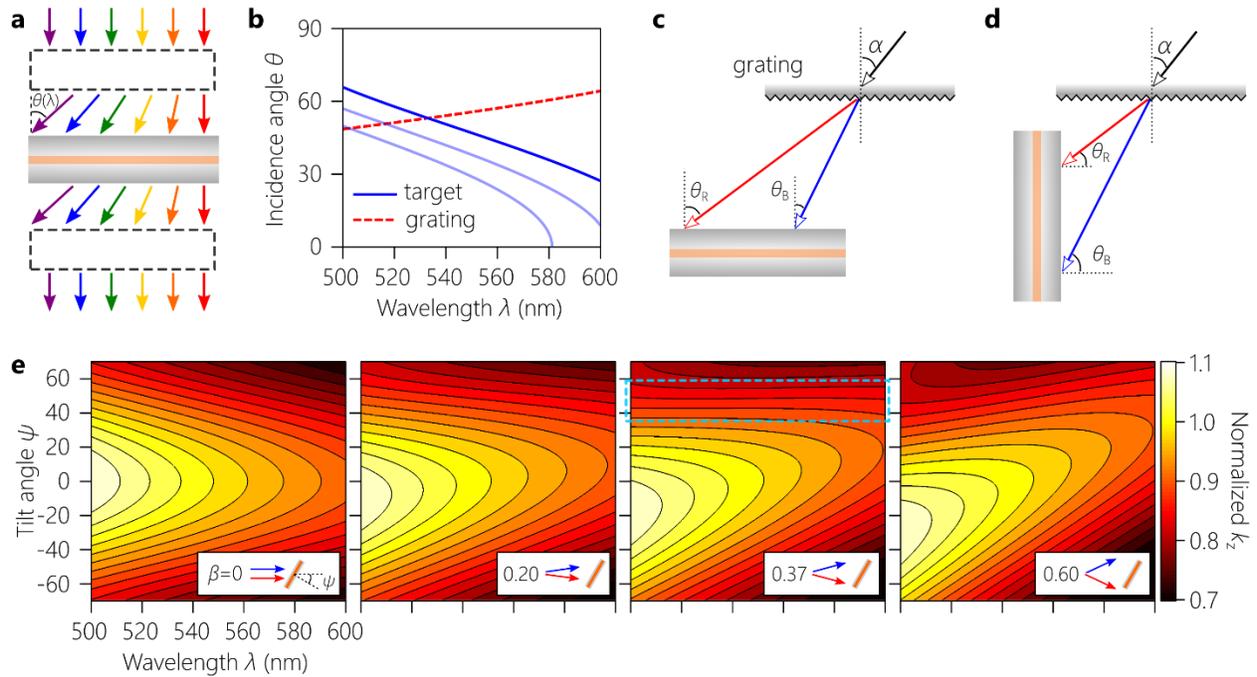

**Figure 2 | Concept of angularly multiplexed phase-matching to produce an achromatic resonance.**
**a**, Using an appropriate 'black-box' system correlating $\lambda$ with $\theta$ (as in Fig. 1c), a planar micro-cavity is rendered transparent. The inverse of this system is placed after the cavity to restore the original beam. **b**, The solid curves are target correlations between $\lambda$ and $\theta$ that help de-slant different resonant mode-orders in a planar micro-cavity (corresponding to the highlighted resonances in Fig. 3d). The dashed curve corresponds to the correlation imparted to a collimated broadband beam centered at $\lambda_c = 550$ nm that is incident normally on a planar surface grating having 1800 lines/mm. **c,d**, Angular diffraction resulting from a planar surface grating (**c**) parallel and (**d**) normal to the plane of a cavity. The former configuration produces the grating curve in (**b**) when $\alpha = 0$. **e**, Calculated $k_z(\lambda, \psi)$ normalized with respect to $k_c = n\frac{2\pi}{\lambda_c}$ in a planar layer of index $n = 1.5$. The highlighted region in the third panel where $\beta = 0.37$ °/nm is independent of $\lambda$ and thus can support achromatic resonances. Insets in each panel in (**e**) depict the corresponding configurations of broadband light incident on the planar layer.

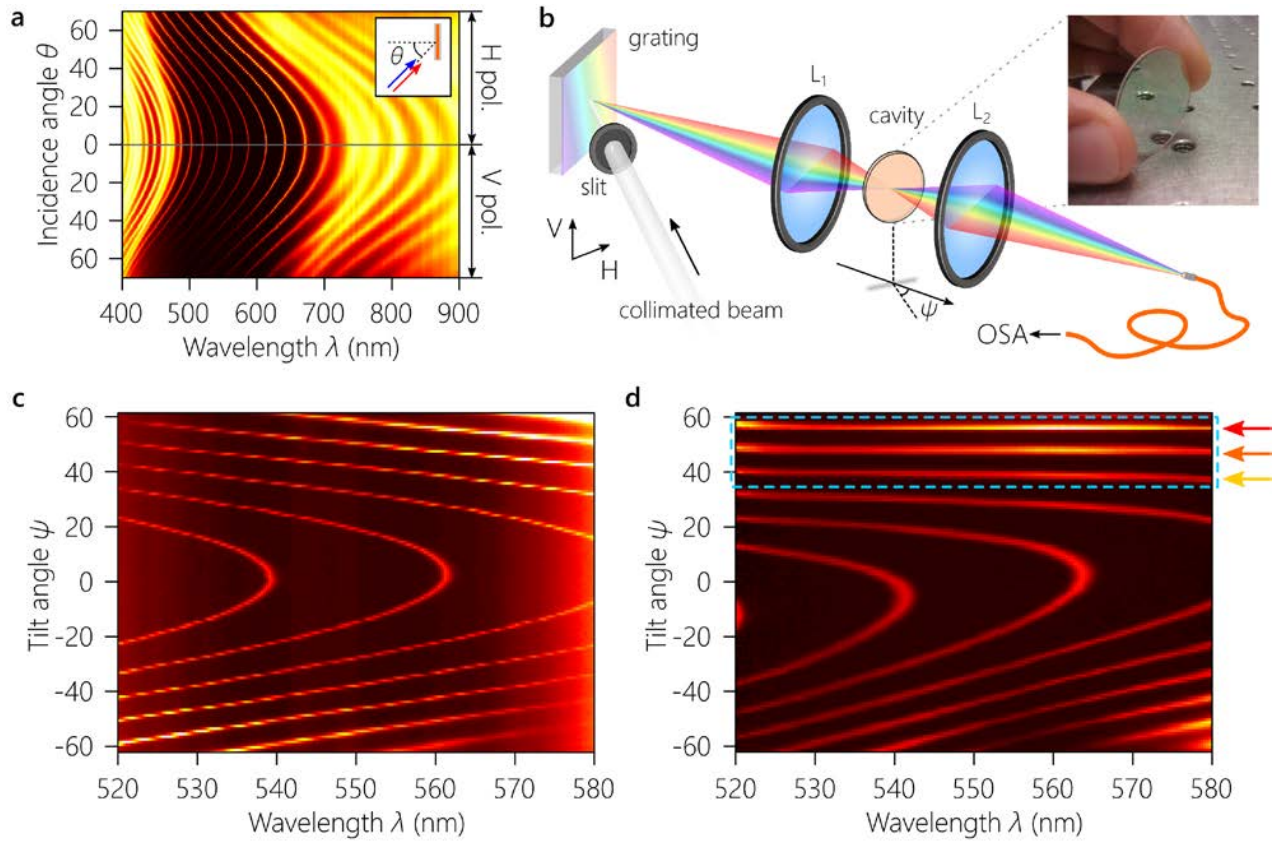

**Figure 3 | Experimental demonstration of achromatic resonances in a Fabry-Pérot micro-cavity. a**, Measured spectral transmission of collimated light through the cavity with angle of incidence $\theta$ for both polarizations. The transmission is symmetric in $\theta$ for TE (H: horizontal) and TM (V: vertical) polarizations, so measurements for only positive $\theta$ are plotted. Inset is a schematic of the configuration. **b**, Experimental setup. $L_1$ and $L_2$ are lenses, OSA: optical spectrum analyzer; see main text and supplementary for details. Inset is a photograph of the resonator showing strong reflectivity in the visible (cavity sample diameter is 25 mm). **c**, Measured spectral transmission through the cavity with tilt angle $\psi$ when the focal length of $L_1$ is $f = 50$ mm. Although the blue-shift with $\psi$ has been enhanced with respect to that in (**a**), the resonances have *not* been completely de-slanted and thus remain *chromatic*. **d**, Measured spectral transmission as in (**c**), except that the focal length of $L_1$ is $f = 25$ mm. The highlighted resonances are completely de-slanted and are now *achromatic* over the bandwidth shown. The grating and lens $L_1$ realize the target $\theta(\lambda)$ correlation functions in Fig. 2b at the three highlighted tilt angles. The measurements in (**c**) and (**d**) are obtained for the H polarization in 1° steps for $\psi$.

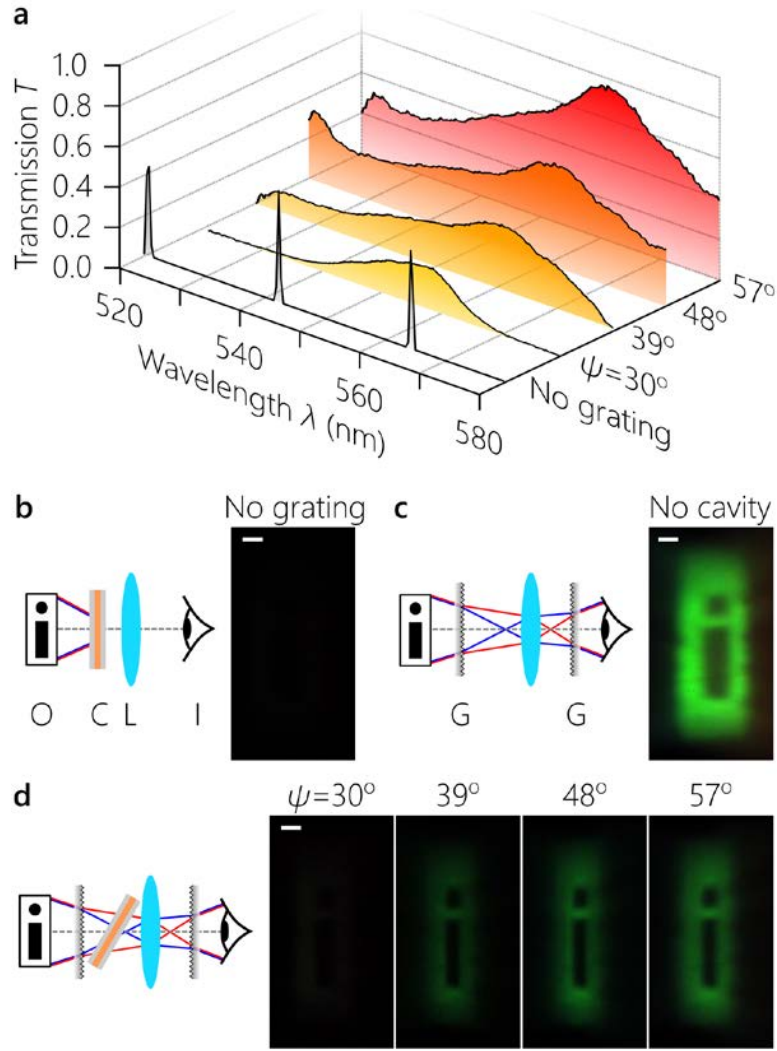

**Figure 4 | Imaging through achromatic cavity resonances. a**, Measured spectral transmission for the achromatic resonances highlighted in Fig. 3d. We also plot the measured bare-cavity normal-incidence transmission (in absence of gratings). **b**, Imaging an object through the cavity. The image is not visible because light is transmitted only through narrow resonances – corresponding to the 'No grating' condition in (**a**). **c**, Imaging an object in absence of the cavity. Here all the source spectrum contributes to the image. **d**, Imaging an object through achromatic resonances at the angular settings depicted in (**a**). On the left of the panels (**b**)-(**d**) we illustrate the imaging configuration and on the right we display the CCD image; scale bars are all 250 μm. O: object plane, C: cavity, L: imaging lens, I: image plane (location of the CCD camera), G: grating; see Supplementary for details. The CCD camera gain is held fixed throughout the measurements.

# Omni-resonant optical micro-cavity

Soroush Shabahang[1], H. Esat Kondakci[1], Massimo L. Villinger[1], Joshua Perlstein[2], Ahmed El Halawany[1], and Ayman F. Abouraddy[1,2*]

[1]CREOL, The College of Optics & Photonics, University of Central Florida, Orlando, FL 32816, USA

[2]Materials Science and Engineering Department, College of Engineering and Computer Science, University of Central Florida, Orlando, FL 32816, USA

*raddy@creol.ucf.edu

# Supporting Information

## 1. Structure of the Fabry-Pérot cavity

The planar Fabry-Pérot (FP) cavity used in the main text is composed of two symmetric 5 bilayer Bragg mirrors enclosing a 4-µm-thick $SiO_2$ dielectric spacer (on a BK7 substrate). The overall structure thus has the layered form:

Incidence → Air – $(HL)_5$ – $SiO_2$ – $(LH)_5$ – BK7.

Here each bilayer (HL) consists of a high-index (H) and low-index (L) material, which are $TiO_2$ and $SiO_2$, respectively. The measured refractive indices for $TiO_2$ and $SiO_2$ at representative wavelengths of interest are given in Table 1 and Table 2 (provided by Blue Ridge Optics, LLC). The $TiO_2$ films were formed by evaporating $Ti_2O_3$ source material under $O_2$ partial pressure. Using these values in Tables S1 and S2, we calculated the spectral response of the full cavity (Figs. S2,S3). The spectral simulations employ the transfer matrix method at a resolution of $(\pi/2)/500$.

**Table S1. Refractive index of $TiO_2$ (produced from $Ti_2O_3$)**

| $\lambda$ | $n$ |
|---|---|
| 470 | 2.1 |
| 510 | 2.09 |
| 550 | 2.085 |
| 590 | 2.08 |
| 670 | 2.075 |

**Table S2. Refractive index of $SiO_2$**

| $\lambda$ | $n$ |
|---|---|
| 450 | 1.4793 |
| 500 | 1.4780 |
| 550 | 1.4772 |
| 570 | 1.4769 |
| 600 | 1.4766 |

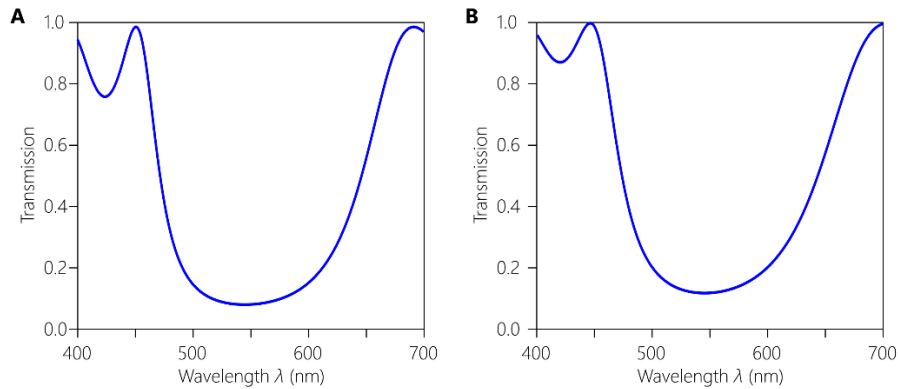

**Fig. S1**. Spectral transmission through the 5-bilayer Bragg mirror at normal incidence. (**A**) The mirror is surrounded by glass on one side and air from the other. (**B**) Mirror is surrounded by glass on both sides.

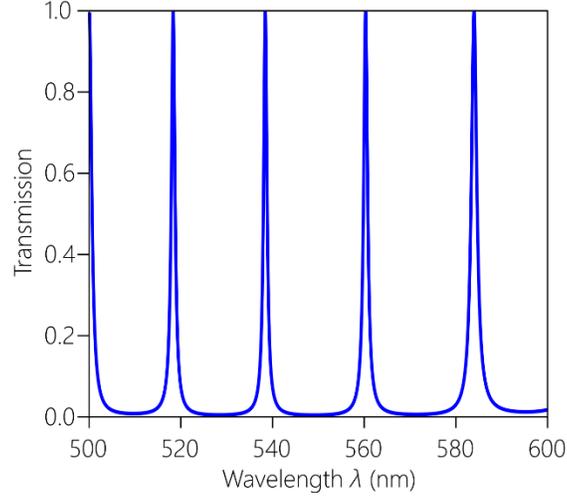

**Fig. S2**. Spectral transmission through the FP cavity at normal incidence. The free spectral range is ~ 25 nm.

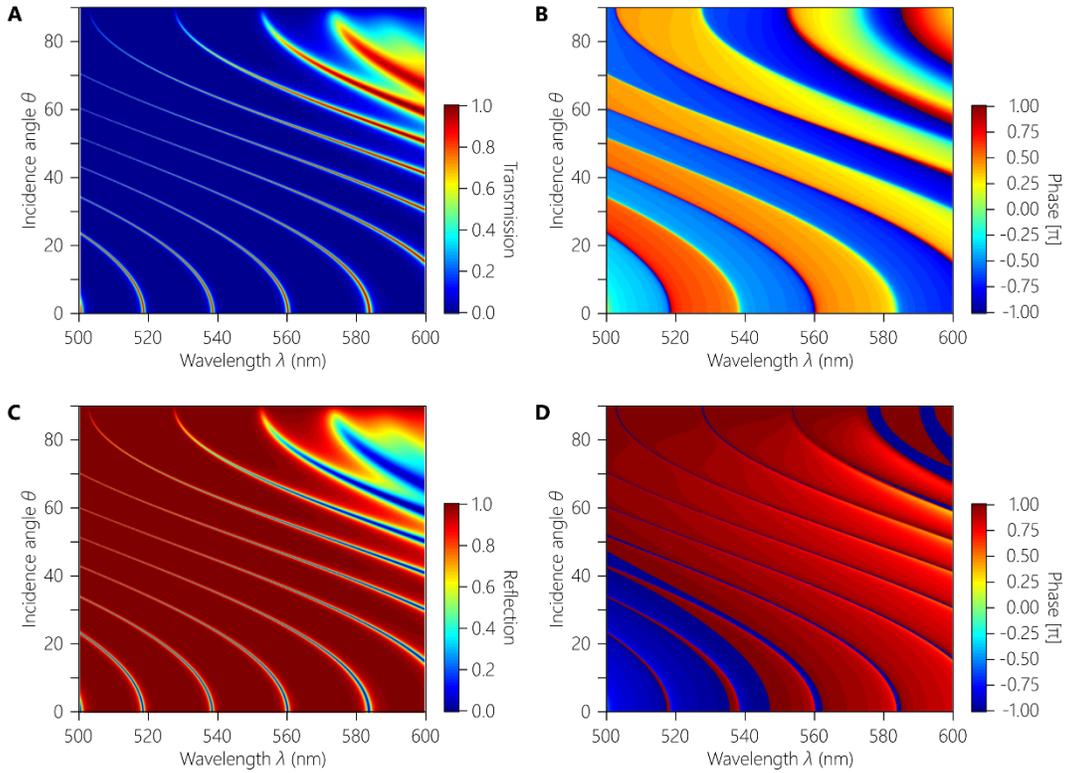

**Fig. S3**. Spectral-angular transmission through and reflection from the FP cavity for incidence of the TE polarization from air. (**A**) Transmittance amplitude and (**B**) phase as a function of the angle of incidence in the spectral range of the Bragg mirror bandgap; see Fig. S1. (**C**) Reflectance amplitude and (**D**) phase. Compare (**A**) to the experimental results reported in Fig. 3A of the main text.

## S2. Simulation of the achromatic resonances

In this Section, we calculate the transmission characteristics of the FP cavity when it is inserted into a setup that induces achromatic resonances. In particular, we simulate the effect of the grating and lens $L_1$ placed in the path of a collimated broadband beam, as shown in Fig. 3B of the main text.

We assume an ideal grating with TE or TM polarized collimated light directed at an incidence angle $\alpha = 50°$ with respect to the normal to the grating. See Fig. S4 for a schematic of the setup that highlights the definition of the relevant angles for our analysis. The angularly dispersed light from the grating is then directed to the sample through the lens $L_1$. We assume that 550 nm is the central wavelength and take it to define the optical axis. The tilt angle of the sample $\psi$ is measured with respect to this optical axis. We define the angle $\gamma(\lambda)$, which is the diffraction angle with respect to the grating normal. The central wavelength $\lambda_c = 550$ nm is diffracted at $\gamma_o = \gamma(\lambda_c = 550 \text{ nm})$ and coincides with the optical axis. The angle any wavelength $\lambda$ makes with respect to this optical axis is $\gamma(\lambda) - \gamma_o$. This angle is boosted via the lens $L_1$ by a ratio $\frac{d_1}{d_2}$, where $d_1$ and $d_2$ are the distances from the grating to $L_1$ and from $L_1$ to the cavity, respectively. The incidence angle made by a wavelength $\lambda$ after the lens with respect to the optical axis is thus:

$$\varphi(\lambda) = \tan^{-1}\left\{\frac{d_1}{d_2}\tan(\gamma(\lambda) - \gamma_o)\right\},$$

with $\varphi_o = \varphi(\lambda_c = 550 \text{ nm}) = 0$. The distances $d_1$ and $d_2$ are selected such that the illuminated spot on the grating is imaged onto the cavity. If the focal length of $L_1$ is $f$, then $d_2 = \frac{f d_1}{f - d_1}$. When the cavity is oriented such that it is perpendicular to the optical axis, the angle of incidence of each wavelength is $\varphi(\lambda)$. Upon tilting the cavity by $\psi$, the angle of incidence with respect to the normal to the cavity is $\theta(\lambda) = \varphi(\lambda) + \psi$.

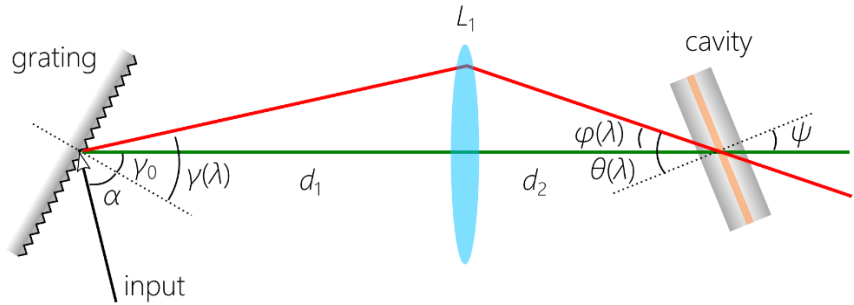

**Fig. S4**. Schematic of the configuration of the grating and cavity to highlight the definitions of the various relevant angles. $\alpha$ and $\gamma(\lambda)$ are measured with respect to the normal to the grating. The optical axis (shown in green) coincides with $\gamma_o = \gamma(\lambda_c = 550 \text{ nm})$. $\varphi(\lambda)$ is measured with respect to the optical axis, while $\theta(\lambda)$ is measured from the normal to the cavity: $\theta(\lambda) = \varphi(\lambda) + \psi$, where $\psi$ is the tilt angle of the cavity with respect to the optical axis.

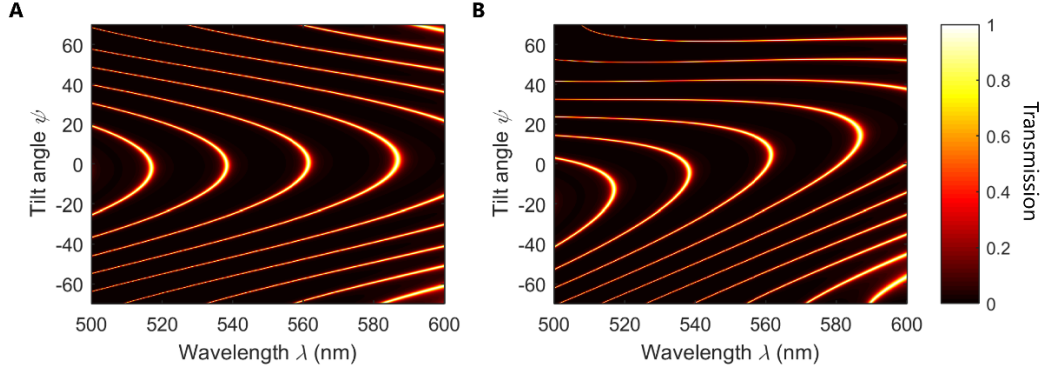

**Fig. S5.** Spectral transmission through an achromatic resonator for TE polarization while varying the cavity tilt angle $\psi$. (**A**) The focal length of L$_1$ is $f = 50$ mm and $d_2 = 8$ cm. (**B**) Same as (**A**) for $f = 25$ mm and $d_2 = 12$ cm. Compare (**A**) and (**B**) to the measurements in Fig. 3C and Fig. 3D in the main text.

With these parameters, we calculate the transmission through the sample using the transfer matrix method for both TE (Fig. S6) and TM (Fig. S7) polarizations. We carry out the calculations for two values of the focal length, $f = 50$ mm (Fig. S5A) and $f = 25$ mm (Fig. S5B) corresponding to the values used in our experiment. The calculations in Fig. S6A and Fig. S6B are to be compared to the measurements reported in the main text in Fig. 3C and Fig. 3D, respectively.

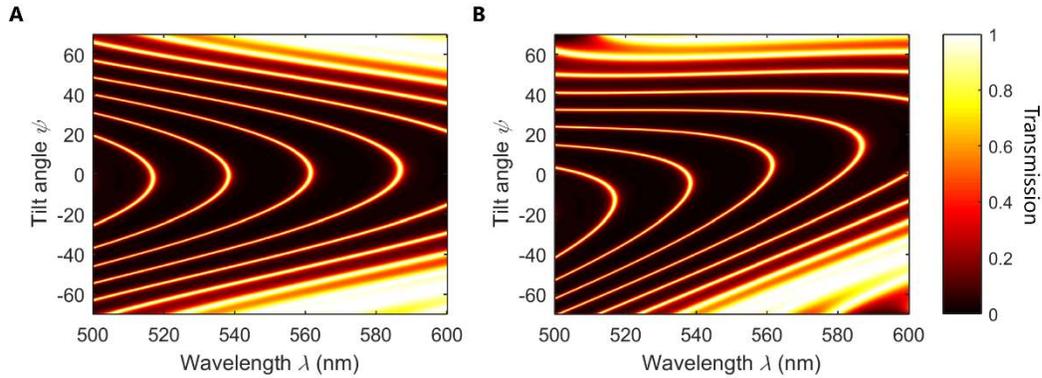

**Fig. S6.** Same as Fig. S5 for TM polarization.

### S3. Experimental setup

**Details of the experimental setup for measuring spectral transmission**

In the experimental setup shown schematically in Fig. 3B in the main text, white light emitted from a halogen source (Thorlabs, QTH10/M) is spatially filtered by coupling through a 1-m-long multimode fiber (50-µm-diameter) and then collimated via a fiber collimator, followed by a polarizer to control the state of polarization of the beam. A grating (Thorlabs GR25-1850) is placed at ~ 30 cm from the collimator and orientated at 50º with respect to the incident light. Before incidence on the grating, the beam passes through a 1-mm-wide vertical slit. Diffracted light is focused by a lens L$_1$ on the cavity, which is mounted on a rotational stage. The lens is placed 12 cm away from the grating to provide the appropriate angular dispersion for phase-matching of the wave-vector axial component. Light transmitted though the cavity is collimated by a second 25-mm-focal-length lens $L_2$ and then coupled into a multimode fiber using an aspherical lens (15-mm-focal-length) connected to a spectrometer (JAZ, Ocean Optics). The wavelength resolution of the measurement is limited by the multimode fiber to ~ 1 nm.

Light diffracted by the grating is spread horizontally, so transversal displacements of optical components can spectrally shift the resonances. To align the setup for the desired spectral range, care must be taken to ensure that the center wavelength of $\lambda_c = 550$ nm passes through the center of the lenses $L_1$ and $L_2$ and thus defines the optical axis. To maximize the achromatic resonance bandwidth, the focusing lens $L_1$ is first placed in the desired distance from the grating obtained from geometrical optics considerations. The collection aspherical lens and fiber are then aligned to collect the maximum spectral bandwidth. The cavity is then mounted on the rotational stage at the focal point of the lens $L_1$. Although axial displacement of the resonator does not affect the resonances, it can alter the angular distribution of the beam after $L_1$, a feature we use to fine-tune the bandwidth of the resonances.

**Details of the experimental setup for imaging through the cavity**

The cavity appears like a mirror at near normal incidence (Fig. 3B inset in the main text), however it transmits most of the incident light in the achromatic resonance configuration and thus appears transparent. To visually demonstrate the cavity transparent, an object (an opaque letter 'i' on a transparent substrate) is imaged onto a CCD camera through the cavity. We first imaged the object through the cavity alone (Fig. 4C in the main text). The cavity is oriented normally with respect to the beam path and as a result blocks most of the optical power due to very low optical throughput and instead reflects the incident beam (except on resonance). The setup is sketched in Fig. S7.

We next carry out the measurement in the achromatic resonance configuration; see Fig. S8. An H-polarized (TE) collimated white light beam is incident on the object (letter 'i' on a transparent substrate), is angularly dispersed by the reflective diffraction grating, followed by passage through the lens $L_1$ with focal length $f = 25$ mm. The beam passes through the cavity mounted on a rotation stage and is collimated by a second 25-mm-focal-length lens $L_2$. A second reflective grating restores the original beam structure and finally a 10-cm-focal lens $L_3$ images the object onto a CCD camera (Imaging Source, DFK 31BU03). The distances from $L_3$ to the grating and from $L_3$ to the CCD are both ~ 10 cm. As a reference, we carry out the experiment using this setup but in absence of the cavity (Fig. 4B in the main text). The results of the imaging experiment in presence of the cavity are presented in Fig. 4D of the main text. At certain values of $\psi$, the resonance broadens, and when the achromatic resonance condition is met the optical throughput increases.

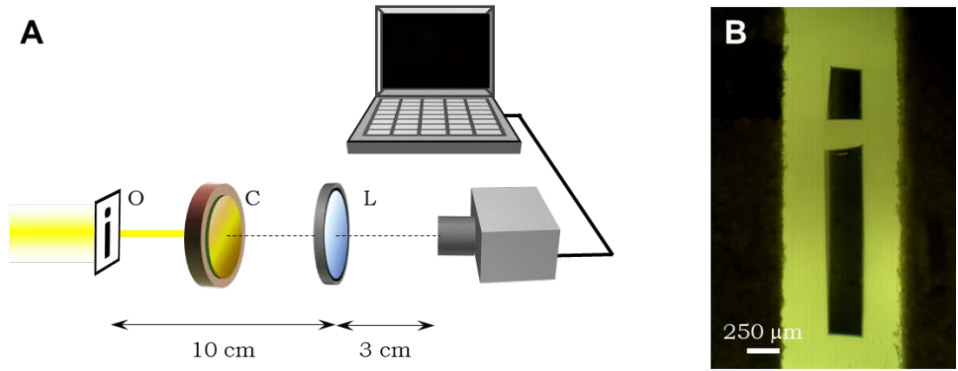

**Fig. S7**. (**A**) Optical setup for imaging an object (letter 'i') through the FP cavity. L: imaging lens; C: FP cavity; O: object plane. (**B**) Optical transmission micrograph of the object.

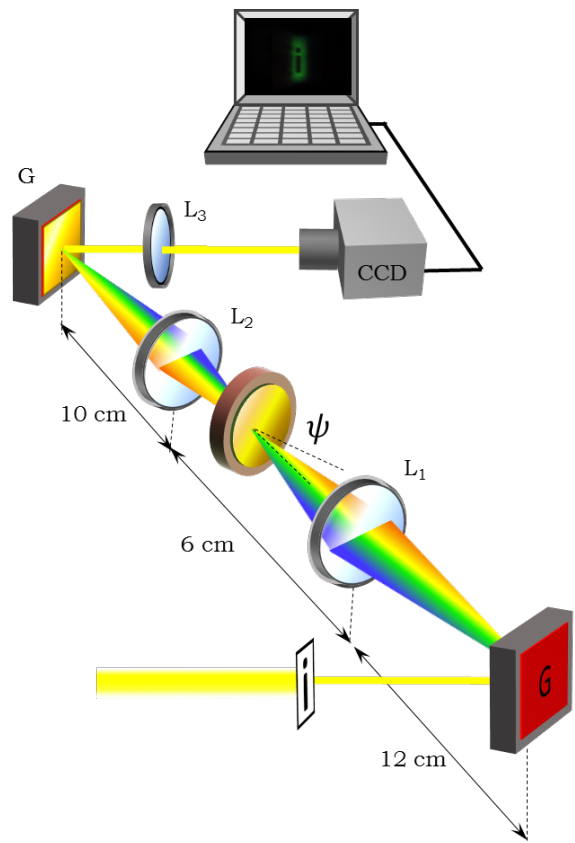

**Fig. S8**. Optical setup for imaging an object (letter 'i') through the FP cavity in a configuration that induces achromatic transmission. G: grating; $L_1$, $L_2$, and $L_3$: lenses.

7